Title:

**Inelastic electron scattering at large angles: the phonon polariton contribution**

Hongbin Yang[1], Paul Zeiger[2], Andrea Konečná[3,4], Lu Han[5], Guangyao Miao[6], Yinong Zhou[7], Yifeng Huang[7], Xingxu Yan[1], Weihua Wang[6], Jiandong Guo[6], Yuefeng Nie[5], Ruqian Wu[7], Jan Rusz[2], Xiaoqing Pan[1,6,8,*]

1. Department of Materials Science and Engineering, University of California, Irvine, CA 92697, USA
2. Department of Physics and Astronomy, Uppsala University, P.O. Box 516, Uppsala 75120, Sweden
3. Institute of Physical Engineering, Brno University of Technology, Brno 61669, Czech Republic
4. Central European Institute of Technology, Brno University of Technology, Brno 62100, Czech Republic
5. National Laboratory of Solid State Microstructures, Jiangsu Key Laboratory of Artificial Functional Materials, College of Engineering and Applied Sciences, Nanjing University, Nanjing, 210093, China.
6. Institute of Physics, Chinese Academy of Sciences, Beijing, 100190, China
7. Department of Physics and Astronomy, University of California, Irvine, CA 92697, USA
8. Irvine Materials Research Institute (IMRI), University of California, Irvine, CA 92697, USA
*Corresponding author: xiaoqinp@uci.edu

**Abstract**

We explore the inelastic electron scattering in $SrTiO_3$, $PbTiO_3$, and SiC in their phonon energy range, challenging the assumption that phonon polaritons are excluded at large angles in high-resolution transmission electron energy-loss spectroscopy. We demonstrate that through multiple scattering, the electron beam can excite both phonons and phonon polaritons, and the relative proportion of each varies depending on the structure factor and scattering angle. Integrating dielectric theory, density functional theory, and multi-slice simulations, we provide a comprehensive framework for understanding these interactions in materials with polar optical phonons.

**Main**

Elementary excitations in solids contain rich information on their electronic, vibrational, and magnetic properties. Many of these excitations, including phonons, plasmons, and magnons, are found in the few-to-hundreds meV energy range. The recent realization of sub-10 meV energy resolution in the scanning transmission electron microscope (STEM) [1] has enabled experimental investigation of such low-energy losses of keV electrons. Thanks to the combined high spatial and energy resolution, distinct vibrational excitations in thin films and nanoparticles [2–7], near defects and interfaces [8–11], have been observed by electron energy-loss spectroscopy (EELS) in the monochromated STEM. However, the interpretation of the inelastic scattering signal has been a non-trivial task. Fast electrons interact strongly with both acoustic and optical phonons in the bulk of a material as well as the long-ranged electric field of surface phonons outside a material, which is one major distinction in the scattering physics between EELS and photon- or neutron-based techniques. The observation of both surface and bulk phonon modes has allowed distinguishing between the so-called dipole and impact scattering [2,12,13].

The dipole scattering is typically associated with the excitations of optically active vibrational modes, which can occur even with a distant electron beam [14,15], and thus the signal becomes delocalized. On the other hand, impact scattering yields the excitation of acoustic or optically inactive modes and yields with a very localized signal [16,17]. The two types of interactions result in different angular distributions of electrons after the scattering events [18,19]. Electrons associated with dipolar scattering receive only a small change in momentum and scattering angle, while the electrons that undergo the impact scattering are found at larger scattering angles. As spatial resolution is essential for probing local phonons near defects and interfaces, there has been great interest in understanding and control of the localization of the acquired inelastic scattering signal [19,20]. It has been shown that by collecting vibrational EELS at large angles, one obtains nanometer spatial resolution from the more localized inelastic signal [12]. Further experiments have demonstrated atomic resolution imaging with phonon scattering even in polar materials [16,21,22]. Thereafter, dark-field (DF) vibrational EELS has emerged as the primary way for probing phonons in the monochromated STEM [10,11,22–24]. The interpretation of DF vibrational EELS has been based on phonon density of states (DOS) calculated from first principles [25] or molecular dynamic simulations [17,26]. This approach can reproduce the approximate peak energy and the overall spectral shape of the loss spectrum. However, phonon polaritons or surface phonon modes are not considered, as they originate from the long-range dipole interaction. This is based on the consensus that DF vibrational EELS signal contains solely bulk phonons from the impact scattering.

Here, we report experimental results that are at odds to the aforementioned understanding and provide a more comprehensive interpretation of vibrational EELS. We perform high-energy resolution STEM-EELS for polar materials in two heterostructures, $PbTiO_3/SrTiO_3$ and FeSe/SiC.

Characteristic spectral features of dipolar phonon polaritons are observed regardless of scattering angle. This indicates that dipolar polariton scattering has a non-negligible contribution in DF vibrational EELS. The observed phenomenon is attribute to multiple scattering of the incident electron beam, which has lost energy to excite phonon polaritons even before penetrating the material. The appropriate interpretation of spectral features thus requires simulations of both phonons and phonon polaritons. By combining the calculations from dielectric theory, density functional theory (DFT), and frequency-resolved frozen phonon multi-slice (FRFPMS) simulations, we provide a quantitative account for the vibrational EELS of SrTiO$_3$, thus showcase a strategy for treating polaritonic scattering in EELS when dealing with polar materials.

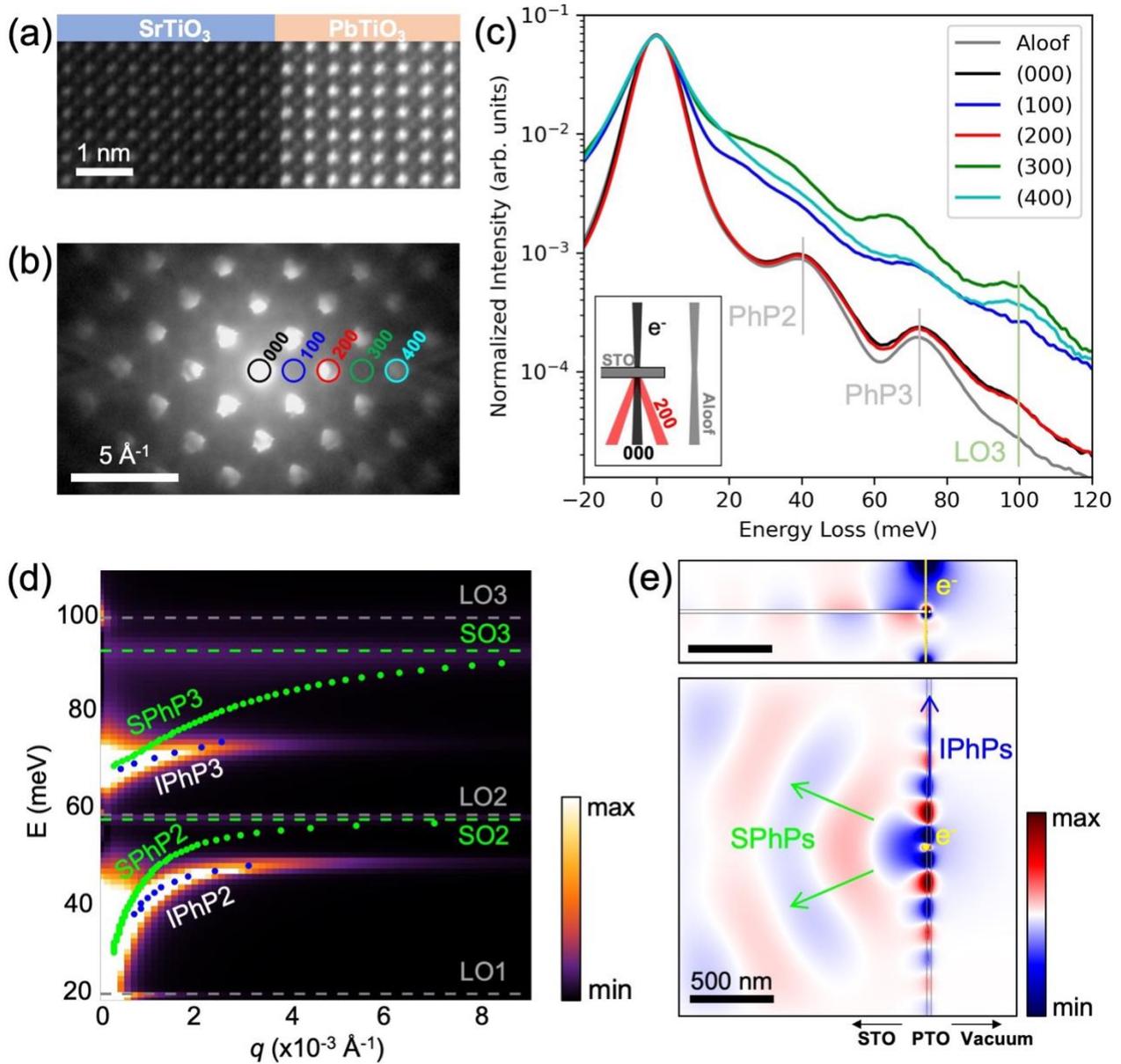

Figure 1. (a) HAADF-STEM image of the PbTiO$_3$-SrTiO$_3$ heterostructure. (b) convergent beam diffraction pattern of SrTiO$_3$ from [010] direction shown in log scale. (c) Raw EEL spectra from the aloof beam, direct beam, and Bragg diffracted beam. The size and locations of the EELS apertures are marked in the diffraction pattern in (b). The inset shows a schematic of the momentum-resolved EELS experiment. (d) Simulated energy-loss probability of electron beam passing through a SrTiO$_3$ slab with an impact parameter (distance of the electron beam from the neighboring PbTiO$_3$ film) b = 10 nm. The LO/SO dashed lines indicate longitudinal/surface optical phonons. The colored dots track dispersions of surface or interface phonon polaritons (marked as SPhP or IPhP). (e) Real-space maps of the total electric field associated with PhP2 at 47 meV and near q=0. The top and bottom panels show the field parallel and perpendicular to the electron beam direction, respectively. The electron beam is indicated by yellow line and dot in the top and bottom panels.

We first show that phonon polaritons are present not only at the Γ point in on-axis EELS, but also at the Γ' points in off-axis EELS. The high-angle annular dark field (HAADF) STEM image in Fig. 1 (a) shows the atomic structure of the $PbTiO_3$-$SrTiO_3$ (PTO-STO) heterostructure in our study. The atomic resolution image is acquired with an Å-sized electron beam, achieved via a large beam convergence angle (33 mrad). Under such probe setting, the in-plane momentum (q) transfer is about 2 Å$^{-1}$, large enough to cover contributions over multiple Brillouin zones of STO. In order to obtain the q-resolved EELS, we instead use a small convergence probe setting (3 mrad), which provides a finely focused beam with 0.4 Å$^{-1}$ q-resolution and ~2.5 nm spatial resolution. The convergent-beam electron diffraction (CBED) pattern of STO acquired with this setting is shown in Fig. 1 (b). We then acquire q-resolved EELS by placing the EELS entrance aperture at different Bragg diffraction spots (N00), where N = 0, 1, 2, 3, 4. Due to the structure factor of STO, the odd diffractions (i.e., (100), (300)) are extremely weak compared to the even ones (i.e., (200), (400)). Diffraction spots represent reciprocal lattice points, regardless of their intensity. Thus, all Bragg angles correspond to the Γ or Γ' points in the reciprocal space.

Monochromated EEL spectra taken at the Γ and Γ' points are shown in Fig. 1 (c). One sees that the (000) and (200) spectra are almost identical and, furthermore, their main loss features in the (000) and (200) spectra also closely resemble those of the aloof spectrum (recorded with the electron beam placed in vacuum). All three spectra have two prominent energy-loss peaks at 44 and 73 meV. As the scattering theory for the direct beam has been well developed in recent years and successfully applied to simulate spectra containing polaritonic energy losses [27], we use it to interpret the energy losses in the (000) spectrum. We simulate the energy loss spectrum of STO with the input of its dielectric function and the sample geometry (Supporting Information). The energy-loss probability of an electron beam passing through a truncated slab of STO near (b = 10 nm) an interface with STO is shown in Fig. 1 (d). We see two branches of strong energy loss within the second and third Reststrahlen band. Within each Reststrahlen band, there exists two types of the phonon polaritons (PhPs): the surface phonon polaritons (SPhPs) which propagate on the top and bottom surface of STO, as well as the interface phonon polaritons (IPhPs) which propagate along the PTO/STO interface. The excitation of these PhPs is responsible for the 44 and 73 meV energy losses. The similarity between the (000), (200), and aloof spectra clearly indicates that they are all dominated by the same PhP energy losses. We note that the PhP energy depends not only on the dielectric properties of the material but also on its shape, size, and surrounding medium. The total electric field of the PhPs and the passing electron beam is shown in Fig. 1(e). For the polaritonic excitations (PhP2 and PhP3), the electric field spreads broadly in real space, which is the reason for the delocalization of dipolar scattering.

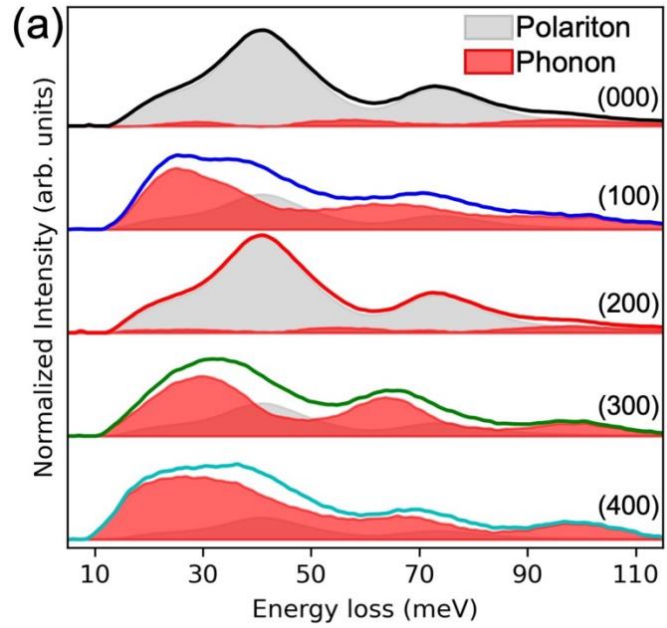

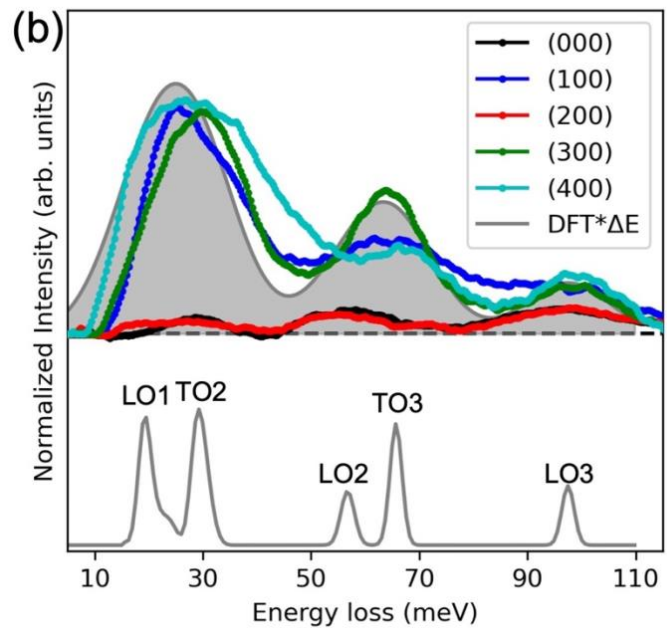

Figure 2. (a) Background-subtracted EEL spectra of STO, together with fitting of the polariton (gray) and phonon (red) components. The spectra are normalized by the integrated intensity and offset by a constant. (b) Phonon components from each Bragg spot compared with DFT-calculated Γ-point phonon DOS. The DFT phonon DOS convolved with experimental energy resolution is shown in gray area.

Having identified the origin of the 44 and 73 meV peaks as phonon polaritons, we may now remove their contribution to get a clear view of the phonon component in each spectrum. By fitting the pure polariton signal (the background-subtracted aloof EELS) to each EEL spectrum, we obtain the percentage of polariton scattering in the total energy loss and then subtract the PhPs from the total energy loss. The resulting polariton and phonon components are shown in Fig. 2(a) as gray and red shaded regions, along with the background-subtracted EEL spectra. The phonon components (red shaded regions in Fig. 2(a)) are gathered in Fig. 2(b) for the comparison with theory.

To show the usefulness of such clean separation of PhP and local phonon contributions in EELS, we calculate the phonon dispersion of STO by the density functional perturbation theory [28]. Our first-principles calculations are performed with the projector-augmented wave pseudopotentials [29,30] and the generalized gradient approximation of Perdew−Burke−Ernzerhof functional [31] using the Vienna Ab initio Simulation Package [32]. An energy cutoff of 450 eV and an 8×8×8 Monkhorst−Pack k-point grid are used [33]. The structure is optimized until the atomic forces are smaller than 0.001 eV/Å. The phonon density of states (DOS) of STO is presented in the bottom panel of Fig. 2(b). Note that our experimental data with the small convergence probe setting are mainly from small q phonons. The integration for phonon DOS is thus performed in the vicinity around the Γ point for q $\leq$ 0.4 Å$^{-1}$ in the Γ-X-M plane. This matches the experimental q resolution and reflect the fact that q transfer associated with phonon energy loss occurs primarily in the plane perpendicular to the beam trajectory. Due to the energy resolution in our EELS experiments, we exclude the phonon modes below 12 meV (acoustic phonons and the soft optical phonon) in the phonon DOS from comparison with experiments. The excitation efficiency of phonons by electron beam may vary due to the character of the vibrational mode. Therefore, we also perform FRFPMS simulations for STO. The results show that both transverse and longitudinal optical (TO and LO) phonons of STO may be excited by the electron beam near the Γ point (SI).

Figure 2(b) shows a detailed comparison between experiment and theory for phonons of STO. The convolved DFT phonon DOS at the Γ point has three main peaks at about 24, 64, and 97 meV originating from the TO and LO phonons. Theory also resolves the EELS peaks at 25 and 64 meV to two phonon modes, i.e., LO1+TO2 and LO2+TO3, respectively. We see that the main features in the (100), (300), and (400) phonon spectra are in good agreement with the convolved DFT phonon DOS. On the other hand, the main signal in the (000) and (200) spectra arise primarily from the LO2 and LO3 phonons. We believe these variations are due to the coupling between photon and TO phonons, i.e., the excitation of the phonon polaritons. For example, when the PhPs are strongly excited at strong Bragg diffractions, the energy loss from the TO phonon excitation is reduced. The observed modulation of TO phonon intensity may even indicate the polariton and phonon scattering are competing loss events, favoring one type of scattering results in a suppression of the other. The weak signal or even absence of Γ-point phonon has also been observed for non-polar materials such as graphite and diamond [34]. The discrepancies near 30 meV and below are not analyzed in detail due to our experimental energy resolution, which makes separating the energy loss signal and the elastic background difficult.

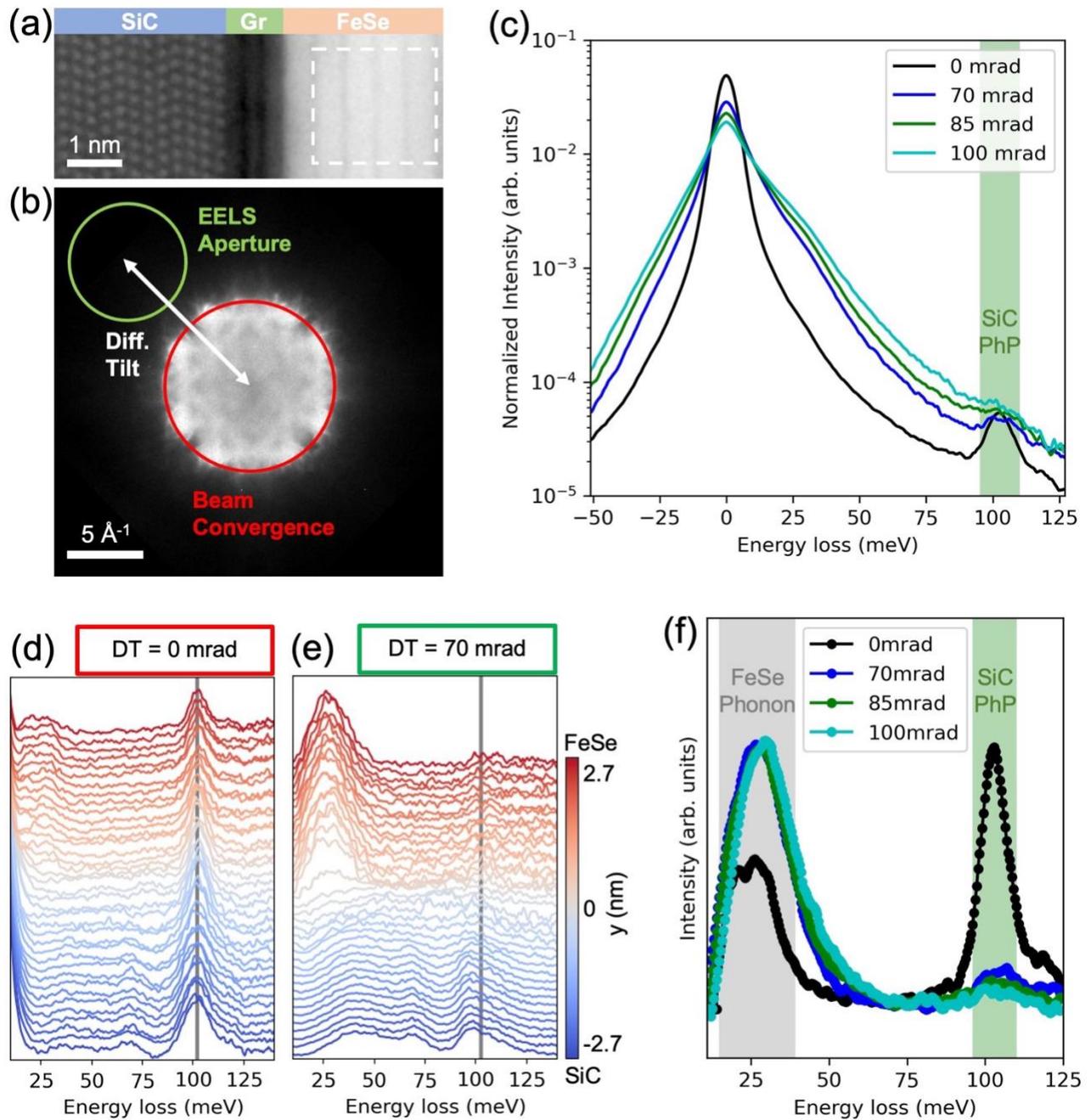

Figure 3. (a) HAADF-STEM image of the FeSe/SiC heterostructure. (b) Schematic of the HAADF-EELS experimental conditions, shown on a 33 mrad convergent beam diffraction pattern of SiC. (c) Raw EEL spectra of FeSe with increasing diffraction tilt (DT). The region for EELS acquisition is marked by the dashed rectangle in (a). (d) On-axis (bright field) and (e) off-axis (dark field) EELS across FeSe/SiC interface, taken at DT angle labeled above. The spectra are plotted with a constant vertical offset of 1.3Å. The line color indicates the acquisition locations. (f) Background-subtracted EELS of FeSe taken with increasing DT angle. The spectra are normalized by the integrated intensity. The EELS acquisition region is marked by the dashed rectangle in (a).

The EEL spectra we have shown so far in Fig. 1 and 2 cover diffraction angles up to 48 mrad / momentum transfer $\Delta q = 6.0$ Å$^{-1}$, which corresponds to medium-angle DF STEM imaging condition. Vibrational EELS with a Å-sized electron beam often requires collecting EELS signal at even larger angles, i.e., in the high-angle annular dark field detection range. Under such conditions, the beam convergence is much larger, thus larger diffraction tilt (DT) angles are required for DF-EELS. As diffraction intensity falls off quickly with the scattering angle, one may expect that the PhP signal will diminish in high-angle dark field (HADF)-EELS. However, spectral evidence for this is still lacking, because the current understanding that small q excitations do not contribute at large angles is inferred from the atomic resolution contrast observed in phonon signal imaging. Due to the overlapping phonon and phonon polariton signal, it is very challenging to separate their respective spectral features in EELS and determine their angular dependence. To quantify the amount of PhPs at high angles experimentally, we study a heterostructure consisting of both light and heavy elements on either side of the interface. An example of such structure is FeSe/SiC interface, where the SiC substrate produces PhPs near 100meV, much higher in energy than the phonons in FeSe. Phonon polaritons are not expected from FeSe as it is a conductor [35]. Taken together, the heterostructure allows minimum energy overlap between the phonons and PhPs compared to those originating from one material.

A HAADF-STEM image of the FeSe/SiC heterostructure is shown in Fig. 3(a), where we observe SiC with graphitized surface and FeSe layers growing on top. We acquire EELS across the interface with increasing DT angle, which controls the distance between the EELS entrance aperture and the BF disk in the diffraction plane. In on-axis EELS (DT = 0 mrad) in Fig. 3(c) and (d), the dipolar SiC PhP near 103 meV is the most prominent loss feature in both SiC and FeSe. The origin of this peak has also been confirmed in Refs. [8,36]. The SiC PhP signal in FeSe barely changes its intensity as a function of distance for the first 3 nm near the interface, as can be seen from the background-subtracted EELS across the interface in Fig. 3(d) and (e). This is expected as the electric field of dipolar PhP is long-ranged and varies only negligibly on the spatial scale of a few nanometers. With increasing DT, the EEL spectra contain more energy loss from impact phonon scattering for both SiC and FeSe, as visible in the spectra acquired with DT = 70 mrad in Fig. 3(e). The variation of FeSe phonon and SiC PhPs with DT are seen more clearly in Fig. 3(f), where the signal was integrated over a larger area in FeSe. While the SiC PhP loss indeed decreased with increasing DT, it did not entirely disappear. Quite surprisingly, we observe the SiC PhP loss in FeSe at DT up to 100 mrad, which is well within the typical HAADF imaging collection angle. Observing PhPs in HADF-EELS clearly shows that dipole scattering is still present at large angles.

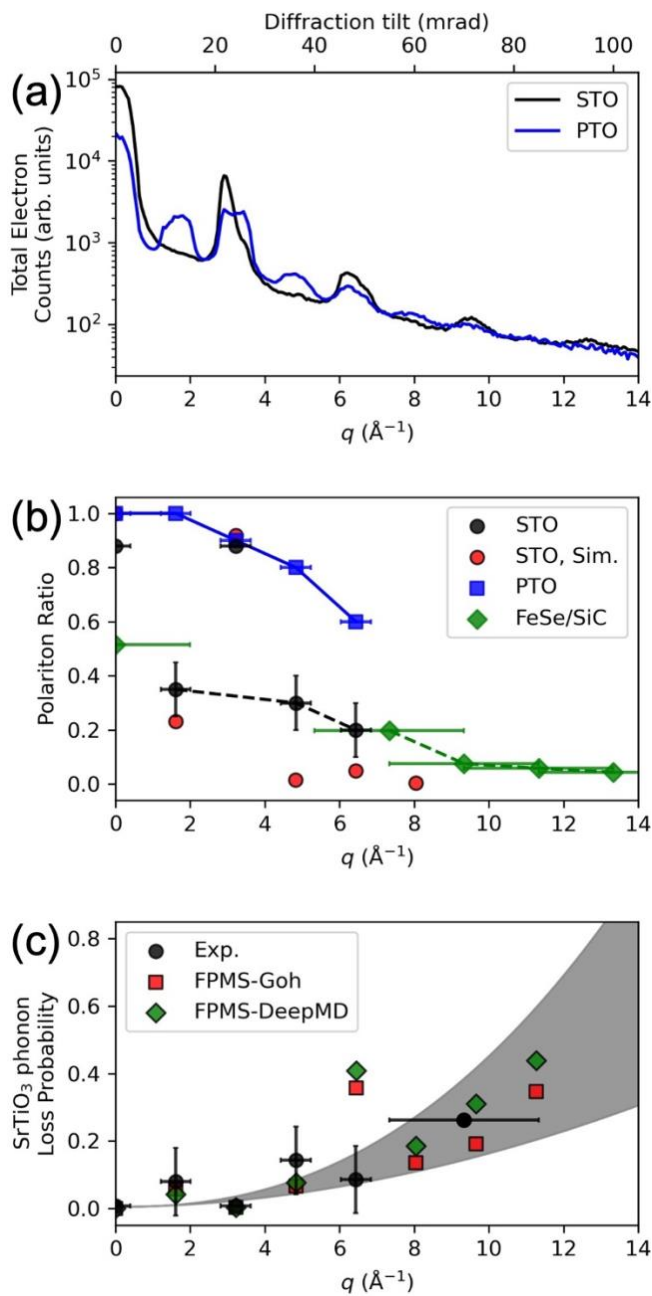

Figure 4. (a) Total electron counts as a function of scattering angle, measured from position averaged CBED pattern of STO and PTO. (b) Polariton ratio in the total energy loss at the Bragg angles for STO and PTO, as well as for FeSe/SiC. (c) Total phonon loss probability of STO from EELS and simulations with different molecular dynamic potentials.

With the results shown above, we further quantify the inelastic scattering probability of fast electrons with a focus on the respective contribution from phonons and phonon polaritons. After the electron beam transmits through a crystalline sample, the majority of electrons is scattered into the Bragg angles, while a small fraction is scattered on phonons and forms the smooth, so-called thermal diffuse scattering background in between [37]. Overall, the total electron counts decrease with the scattering angle, as shown in Fig. 4(a).

The ratio of electron energy loss that arises from polariton excitation (PhP ratio) is plotted in Fig. 4(b). The PhP ratio of the total energy loss is determined by fitting an aloof spectrum to the background subtracted EELS, as demonstrated in Fig. 2(a). We see that the strong Bragg diffractions of STO contain > 90 % PhPs, whereas weak diffraction spots contain much less polariton scattering. We have also measured the $\Gamma$ and $\Gamma$' points vibrational EELS of PTO, a perovskite oxide that does not have weak diffractions. The comparison with STO clearly shows that the PhP ratio indeed depends on the Bragg diffraction intensity. The FeSe/SiC data points represent the ratio between SiC PhPs and the total energy loss measured on the FeSe side of the heterostructure. The measurements are performed at large beam convergence and EELS collection angles. Under this experimental condition, the Bragg disks overlap, which spreads the polariton signal over the whole diffraction plane. Thus, the loss signal consists of a mixture of phonons and polaritons and shows a reduced polariton percentage compared to the q-resolved EELS measured at Bragg angles. We simulate the polariton excitation at the Bragg angles by combining dielectric theory with frozen phonon multi-slice (FPMS) simulations. From the dielectric formalism, we obtain the polariton energy loss probability at $q = 0$, which is then used to calculate polariton loss probability at the Bragg angles. The simulation results in Fig. 4(b) agree relatively well with experiments, both of which show that the PhP ratio is proportional to the Bragg diffraction intensity. At large angles and at weak diffractions, the experimental PhP ratios are larger than the simulated ones. One possible reason for the discrepancies could be an underestimation of phonon energy loss intensity at low energies due to the background subtraction. Nevertheless, the decrease of PhP ratio with the total electron counts is observed in both q-resolved and spatial-resolved conditions and reproduced well in simulations. Importantly, however, the PhP ratio approaches a non-zero value even when q is very large (> 80 mrad).

On the other hand, the electron energy-loss probability that arises from phonon excitation increases with q, as shown in Fig. 4(c). At the $\Gamma$' points with strong Bragg diffractions, the phonon loss probability increases quadratically with q (the gray-shaded region in Fig. 4(c)). The increase in phonon loss probability has also been reproduced by the FRFPMS simulation of the STO phonons. We notice that at some momentum transfer, the phonon loss probabilities are exceptionally large compared with experiments, which can be possibly due to the molecular dynamic potential used, or the underestimated total electron intensity at these angles due to unaccounted strong valence plasmon scattering in our multislice simulation [38].

Finally, we discuss the reason for the presence of dipolar polaritonic scattering in off-axis and DF EELS. The polariton excitation can occur when the electron beam is passing through the material, but also before or after the beam enters the material, as the evanescent field of polaritons can reside well outside of the sample. Thus, the electrons at the Bragg angle undergo multiple elastic scattering in the material and carry the same polaritonic loss events as the direct beam. In our simulations, we notice that the PhP fraction barely changes with how the beam is propagated inside the material (either multi or single-slice) (SI). This is because the majority of the interaction with optically active phonons occurs outside the sample, and even quite far away from the sample surfaces (SI). Therefore, the interior of the material only adds a negligible portion of the total polariton loss. This mechanism also applies to large scattering angles, where the Bragg electron counts are on the same level as the diffuse scattering background. The scattering process will not be altered by tilting the electron beam either before or after the specimen in vibrational EELS experiments. Thus, having a mixed dipole and impact scattering is an intrinsic nature for the energy loss of fast electrons when polar materials are involved. The multiple scattering process observed through transmission EELS is likely also applicable to low-energy reflection EELS [39].

The observation of PhPs in HADF-EELS contrasts with the conclusions made in previous studies, where dipolar scatterings have been thought to be excluded in HAADF imaging. Although the presence of dipolar polaritons may not obscure atomic resolution imaging via the phonon loss signal, it can produce noticeable spectral features in vibrational EELS that require considering both types of scatterings. For example, due to the extended field of STO PhPs, the vibrational EELS of PTO measured in the PTO/STO heterostructure is substantially different from that in pure PTO. Our finding can also help resolve the discrepancies between theory and vibrational EELS experiments in the past. Including PhPs in q-resolved vibrational EELS at the Γ points of BN will provide a more accurate interpretation to the experimental data [25,40–42]. Although our observations shown here are based on phonon polaritons, the phenomenon and scattering physics also applies to other long-ranged excitations, such as plasmon polaritons [43]. Furthermore, our work shows the importance of considering diffraction conditions in future vibrational EELS experiments. The diffraction conditions may vary with the material structure factor, which can cause changes in polariton intensity across interfaces of different materials. Our results also provide guidelines for studying the coupling between phonons and electrons and plasmons, as the couplings are often particularly strong near the $\Gamma$ point in polar materials.

In conclusion, our results provide clear evidence that polaritons in polar materials have a significant contribution in EELS regardless of scattering angle. We demonstrate that integrating theories for elastic and low-loss inelastic electron-matter interaction can provide a more complete interpretation of vibrational EELS for polar materials. Using $SrTiO_3$ as a model system, we analyze the phonon and phonon polariton modes in detail and show that the excitation of polaritons competes with phonon scattering and influences the observed spectrum. We have further quantified the angular distribution of electrons inelastically scattered in several materials and close to their interfaces and revealed the dependence of the polaritonic scattering contribution on Bragg diffraction intensity. This study underscores the necessity of accounting for polariton scattering to fully understanding of vibrational EELS, which is a critical tool for exploring vibrational properties at the atomic level.

**Acknowledgments**
This work was supported by the Department of Energy, Office of Basic Energy Sciences, Division of Materials Sciences and Engineering (DE-SC0014430). HY and XP acknowledge the use of facilities and instrumentation at the UC Irvine Materials Research Institute (IMRI) supported in part by the National Science Foundation through the Materials Research Science and Engineering Center program (DMR-2011967). PZ and JR acknowledge financial support of Swedish Research Council (proj. no. 2021-03848), Olle Engkvist's foundation, and Knut and Alice Wallenberg's foundation. Part of the simulations was enabled by resources provided by the National Academic Infrastructure for Supercomputing in Sweden (NAISS) at the NSC centre partially funded by the Swedish Research Council through grant agreement no. 2022-06725. AK acknowledges the support of the Czech Science Foundation GACR under the Junior Star grant No. 23-05119M. YZ and RW were supported by the U.S. DOE, Office of Science (Grant No. DE-FG02- 05ER46237).

**References**

[1]   O. L. Krivanek et al., *Vibrational Spectroscopy in the Electron Microscope*, Nature **514**, 209 (2014).
[2]   M. J. Lagos, A. Trügler, U. Hohenester, and P. E. Batson, *Mapping Vibrational Surface and Bulk Modes in a Single Nanocube*, Nature **543**, 529 (2017).
[3]   A. A. Govyadinov et al., *Probing Low-Energy Hyperbolic Polaritons in van Der Waals Crystals with an Electron Microscope*, Nat Commun **8**, (2017).
[4]   K. Venkatraman, P. Rez, K. March, and P. A. Crozier, *The Influence of Surfaces and Interfaces on High Spatial Resolution Vibrational EELS from SiO2*, Microscopy **67**, i14 (2018).
[5]   N. Li et al., *Direct Observation of Highly Confined Phonon Polaritons in Suspended Monolayer Hexagonal Boron Nitride*, Nat Mater **20**, 43 (2021).
[6]   H. Lourenço-Martins and M. Kociak, *Vibrational Surface Electron-Energy-Loss Spectroscopy Probes Confined Surface-Phonon Modes*, Phys Rev X **7**, 41059 (2017).
[7]   M. J. Lagos, A. Trügler, V. Amarasinghe, L. C. Feldman, U. Hohenester, and P. E. Batson, *Excitation of Long-Wavelength Surface Optical Vibrational Modes in Films, Cubes and Film/Cube Composite System Using an Atom-Sized Electron Beam*, Microscopy **67**, i3 (2018).
[8]   X. Yan et al., *Single-Defect Phonons Imaged by Electron Microscopy*, Nature **589**, 65 (2021).
[9]   R. Qi et al., *Measuring Phonon Dispersion at an Interface*, Nature **599**, 399 (2021).
[10]  Y. Li, R. Qi, and R. Shi, *Atomic-Scale Probing of Hetero-Interface Phonon in Nitride Semiconductor*, Proc Natl Acad Sci U S A **119**, (2022).
[11]  E. R. Hoglund et al., *Emergent Interface Vibrational Structure of Oxide Superlattices*, Nature **601**, 556 (2022).
[12]  C. Dwyer, T. Aoki, P. Rez, S. L. Y. Chang, T. C. Lovejoy, and O. L. Krivanek, *Electron-Beam Mapping of Vibrational Modes with Nanometer Spatial Resolution*, Phys Rev Lett **117**, (2016).
[13]  U. Hohenester, A. Trügler, P. E. Batson, and M. J. Lagos, *Inelastic Vibrational Bulk and Surface Losses of Swift Electrons in Ionic Nanostructures*, Phys Rev B **97**, 165418 (2018).
[14]  P. Rez, T. Aoki, K. March, D. Gur, O. L. Krivanek, N. Dellby, T. C. Lovejoy, S. G. Wolf, and H. Cohen, *Damage-Free Vibrational Spectroscopy of Biological Materials in the Electron Microscope*, Nat Commun **7**, (2016).
[15]  J. A. Hachtel, J. Huang, I. Popovs, S. Jansone-Popova, J. K. Keum, J. Jakowski, T. C. Lovejoy, N. Dellby, O. L. Krivanek, and J. Carlos Idrobo, Identification of Site-Specific Isotopic Labels by Vibrational Spectroscopy in the Electron Microscope, 2019.
[16]  K. Venkatraman, B. D. A. Levin, K. March, P. Rez, and P. A. Crozier, *Vibrational Spectroscopy at Atomic Resolution with Electron Impact Scattering*, Nature Physics.
[17]  P. M. Zeiger and J. Rusz, *Efficient and Versatile Model for Vibrational STEM-EELS*, Phys Rev Lett **124**, (2020).

[18] N. R. Lugg, B. D. Forbes, S. D. Findlay, and L. J. Allen, *Atomic Resolution Imaging Using Electron Energy-Loss Phonon Spectroscopy*, Phys Rev B Condens Matter Mater Phys **91**, (2015).

[19] C. Dwyer, *Prospects of Spatial Resolution in Vibrational Electron Energy Loss Spectroscopy: Implications of Dipolar Scattering*, Phys Rev B **96**, (2017).

[20] R. F. Egerton, K. Venkatraman, K. March, and P. A. Crozier, *Properties of Dipole-Mode Vibrational Energy Losses Recorded from a TEM Specimen*, Microscopy and Microanalysis **26**, 1117 (2020).

[21] F. S. Hage, D. M. Kepaptsoglou, Q. M. Ramasse, and L. J. Allen, *Phonon Spectroscopy at Atomic Resolution*, Phys Rev Lett **122**, (2019).

[22] F. S. Hage, G. Radtke, D. M. Kepaptsoglou, M. Lazzeri, and Q. M. Ramasse, *Single-Atom Vibrational Spectroscopy in the Scanning Transmission Electron Microscope*, Science (1979) **367**, 1124 (2020).

[23] R. Senga, Y. C. Lin, S. Morishita, R. Kato, T. Yamada, M. Hasegawa, and K. Suenaga, *Imaging of Isotope Diffusion Using Atomic-Scale Vibrational Spectroscopy*, Nature **603**, 68 (2022).

[24] M. Xu et al., *Single-Atom Vibrational Spectroscopy with Chemical-Bonding Sensitivity*, Nat Mater (2023).

[25] R. J. Nicholls, F. S. Hage, D. G. McCulloch, Q. M. Ramasse, K. Refson, and J. R. Yates, *Theory of Momentum-Resolved Phonon Spectroscopy in the Electron Microscope*, Phys Rev B **99**, (2019).

[26] P. M. Zeiger and J. Rusz, *Frequency-Resolved Frozen Phonon Multislice Method and Its Application to Vibrational Electron Energy Loss Spectroscopy Using Parallel Illumination*, Phys Rev B **104**, (2021).

[27] A. Konečná, K. Venkatraman, K. March, P. A. Crozier, R. Hillenbrand, P. Rez, and J. Aizpurua, *Vibrational Electron Energy Loss Spectroscopy in Truncated Dielectric Slabs*, Phys Rev B **98**, (2018).

[28] S. Baroni, S. De Gironcoli, A. D. Corso, and P. Giannozzi, Phonons and Related Crystal Properties from Density-Functional Perturbation Theory, n.d.

[29] P. E. Blochl, *Projector Augmented-+rave Method*, Phys Rev B **50**, 24 (n.d.).

[30] G. Kresse and D. Joubert, From Ultrasoft Pseudopotentials to the Projector Augmented-Wave Method, n.d.

[31] J. P. Perdew, K. Burke, and M. Ernzerhof, *Generalized Gradient Approximation Made Simple*, Phys Rev Lett (1996).

[32] G. Kresse and J. Furthmüller, *Efficient Iterative Schemes for Ab Initio Total-Energy Calculations Using a Plane-Wave Basis Set*, Phys Rev B **54**, 11169 (1996).

[33] M. Methfessel and A. T. Paxton, *High-Precision Sampling for Brillouin-Zone Integration in Metals*, Phys Rev B **40**, 3616 (1989).

[34] J. Kikkawa, T. Taniguchi, and K. Kimoto, *Nanometric Phonon Spectroscopy for Diamond and Cubic Boron Nitride*, Phys Rev B **104**, (2021).

[35] A. Subedi, L. Zhang, D. J. Singh, and M. H. Du, *Density Functional Study of FeS, FeSe, and FeTe: Electronic Structure, Magnetism, Phonons, and Superconductivity*, Phys Rev B Condens Matter Mater Phys **78**, (2008).

[36] Y. Li, R. Qi, R. Shi, N. Li, and P. Gao, *Manipulation of Surface Phonon Polaritons in SiC Nanorods*, Sci Bull (Beijing) **65**, 820 (2020).

[37] B. D. Forbes, A. V. Martin, S. D. Findlay, A. J. D'Alfonso, and L. J. Allen, *Quantum Mechanical Model for Phonon Excitation in Electron Diffraction and Imaging Using a Born-Oppenheimer Approximation*, Phys Rev B Condens Matter Mater Phys **82**, (2010).

[38] D. A. Muller, B. Edwards, E. J. Kirkland, and J. Silcox, Simulation of Thermal Diffuse Scattering Including a Detailed Phonon Dispersion Curve, 2001.

[39] S. Zhang et al., *Enhanced Superconducting State in FeSe/SrTiO3 by a Dynamic Interfacial Polaron Mechanism*, Phys Rev Lett **122**, (2019).

[40] F. S. Hage, R. J. Nicholls, J. R. Yates, D. G. Mcculloch, T. C. Lovejoy, N. Dellby, O. L. Krivanek, K. Refson, and Q. M. Ramasse, *Nanoscale Momentum-Resolved Vibrational Spectroscopy*, Sci Adv **4**, eaar7495 (2018).

[41] R. Senga, K. Suenaga, P. Barone, S. Morishita, F. Mauri, and T. Pichler, *Position and Momentum Mapping of Vibrations in Graphene Nanostructures*, Nature **573**, 247 (2019).

[42] R. Qi et al., *Four-Dimensional Vibrational Spectroscopy for Nanoscale Mapping of Phonon Dispersion in BN Nanotubes*, Nat Commun **12**, (2021).

[43] H. Yang, E. L. Garfunkel, and P. E. Batson, *Probing Free Carrier Plasmons in Doped Semiconductors Using Spatially Resolved Electron Energy Loss Spectroscopy*, Phys Rev B **102**, (2020).